\documentclass[12pt]{article}

\begin{document}

\renewcommand{\]}{$$%
\par  \noindent  \hspace{-0.4em}}

\newcommand{\rf}[1]{(\ref{#1})}
\newcommand{\rff}[2]{(\ref{#1}\ref{#2})}

\newcommand{\ba}{\begin{array}}
\newcommand{\ea}{\end{array}}

\newcommand{\be}{\begin{equation}}
\newcommand{\ee}{\end{equation}}

\newcommand{\const}{{\rm const}}

\newcommand{\Cl}{{\cal C}}

\newcommand{\e}{{\bf e}}

\newcommand{\m}{\left( \ba{c}}
\newcommand{\ema}{\ea \right)}
\newcommand{\mm}{\left( \ba{cc}}
\newcommand{\miv}{\left( \ba{cccc}}

\newcommand{\scal}[2]{\mbox{$\langle #1 \! \mid #2 \rangle $}}

\newtheorem{prop}{Proposition}
\newtheorem{Th}{Theorem}
\newtheorem{lem}{Lemma}
\newtheorem{rem}{Remark}
\newtheorem{cor}{Corollary}
\newtheorem{Def}{Definition}
\newtheorem{open}{Open problem}

\title{\bf  Why the phase shifts for solitons \\ 
on a vortex filament are so large?\thanks{The
research supported partially by the KBN grant
2 P03B 126 22.}}

\author{\bf Jan L.\ Cie\'sli\'nski\thanks{\footnotesize
e-mail: \tt janek\,@\,alpha.uwb.edu.pl}
\\ {\footnotesize Uniwersytet w Bia\l ymstoku,
Instytut Fizyki Teoretycznej}
\\ {\footnotesize ul.\ Lipowa 41, 15-424
Bia\l ystok, Poland}
}

\date{}

\maketitle

\begin{abstract}
The phase ``jumps'' for solitons interacting on a vortex filament, observed in
experiments, have been unaccounted for since more than twenty years. 
Using explicit formulas describing the interaction of two solitons on
a thin vortex filament in the Localized Induction Approximation we show that an appropriate choice of the parameters of the solitons leads to large phase shifts. 
This result does not depend on the axial flow along the filament.  
\end{abstract}

\vspace{0.3cm}

\noindent {\it PACS Numbers:} 05.45Yv, 02.30.Ik

\noindent {\it Key words and phrases:} vortex filament, localized induction equation, solitons, phase shift

\pagebreak

It is well known that the equation describing the motion of a single thin vortex 
filament in an ideal fluid in the localized induction approximation \cite{Ba} 
is integrable in the sense of the soliton theory
\cite{La}) and admits soliton solutions (known also as kinks or Hasimoto vortices) \cite{Ha}. 
It is interesting that the equation itself has been derived
as early as 1906 by da Rios and then rediscovered few times, see \cite{Ri2}).
However, the experiments
carried out some time later have shown that although some
soliton-like patterns can be observed and their qualitative
behavior agrees with the theory (\cite{HB}) but there are some
discrepancies concerning quantitative details \cite{AK}.
Let us mention the ratio of the group velocity and the phase velocity, the dependence between phase velocity,
angular velocity and
group velocity, and last but not least: the very large ``phase
jump'' during head-on collisions of two solitons
\cite{MHR,MMH}.

Then, the approximation has been improved by taking into
account the
axial flow (the flow inside the vortex core, along
the filament) \cite{FM}. The kinematics of the vortex filament
remained integrable but the equation is now more general
(it contains one additional free parameter corresponding to
the intensity and direction of the axial flow) \cite{FM,KI}:
\be \label{axial}
        {\bf r},_t = \alpha ({\bf r},_s\!\times\!\,{\bf r},_{ss}) + \beta ({\bf r},_{sss} +
        \frac{3}{2} ({\bf r},_{ss})^2 {\bf r},_s) \ ,  \qquad  ({\bf r},_s)^2= 1 \ ,
\ee  \par \noindent
where ${\bf r}={\bf r}(s,t) \in {\bf R}^3$, $t$ is the time and
$s$ is arc length parameter along the vortex filament, finally
$\alpha$ and $\beta$ are assumed be constant ($\alpha > 0$). 
The case $\beta=0$ corresponds to Localized Induction Equation of da Rios, closely associated with the cubic
nonlinear Schr\"odinger equation and continuum Heisenberg
ferromagnt model (see \cite{S-2}).
The case $\alpha=0$ is associated with the modified
Korteweg-de Vries equation. Usually $\alpha, \beta$ are treated as 
parameters of the theoretical 
model and they have to follow from experimental data. 

The exact formulas for multi-soliton solutions can be obtained
using different methods (compare \cite{Ha,FM,KI,KMI}. We 
follow the soliton surfaces approach \cite{S-2,Sym} as most
convenient. The one-soliton solution
is obtained without difficulties \cite{Sym,S-6} but two-soliton
solution looks much more complicated (compare \cite{KMI,LSW-vortex}).
An essential simplification of the two-soliton formula
is given in \cite{Ci-twosol,Ci-nos}.

However, the following statement written more than 16 years ago seems to be still accepted:
{\it ``The shapes of the `Hasimoto soliton' seem adequate to fit experimental data, but the
ratio between group velocity and phase velocity for the wave is a constant (2.0) in LIA,
whereas the experiments show a dependence on the ratio of wavelength to core size. Also
unexplained is a large phase advance on collision of two experimental `solitons'.
An intriguing attempt to go beyond LIA as originally conceived was presented by
Fukumoto \& Miyazaki. Their proposal to include an arbitrary axial flow (\ldots)
leaves the shapes of the solitons unchanged, changes
the velocity of the propagation, but, unfortunately, does nothing for the troublesome
phase jump.''} \cite{AK}.

The main goal of the present paper is to show that the large phase jumps are not longer 
``troublesome'' and are in good agreement with the theory.

In order to describe the interaction of two solitons on a single vortex filament 
 we use the following exact formula for two-soliton solution:
\be \ba{l}  \label{N=2}
{\bf r}_2 = 
\displaystyle \frac{1}{2 K}
   \left( \ba{c} d_1 \left( e^{Q_2} \cos\!P_1^+ +
    e^{-Q_2} \cos\!P_1^- \right)    +
   d_2 \left( e^{Q_1} \cos\!P_2^+ + e^{-Q_1} \cos\!P_2^- \right) \\
    d_1 \left( e^{Q_2} \sin\!P_1^+ +
    e^{-Q_2} \sin\!P_1^- \right)    +
   d_2 \left( e^{Q_1} \sin\!P_2^+ + e^{-Q_1} \sin\!P_2^- \right) \\
   d_+ \sinh\!(Q_1+Q_2) + d_- \sinh\!(Q_1-Q_2) +
    d_0 \sin\!(P_1-P_2) - 2 s K 
\ea  \right)
\ea \ee
where 
\[     \ba{l}
K = \cosh\! Q_1 \cosh\! Q_2 \cosh\! \Delta + \sinh\! Q_1 \sinh\! Q_2 \sinh\! \Delta
   + \sinh\! \Delta \cos\!(P_1-P_2) \ ,  \\[3ex]
P_k^\pm = P_k \pm \delta_k  \quad (k=1,2) \ , 
\\[3ex]
   Q_k = 2 s\ {\rm Im} \lambda_k + 2 t\ {\rm Im} \omega_k + Q_{k0} \ , \\[3ex]
    P_k = - 2 s\ {\rm Re} \lambda_k -  2 t\ {\rm Re} \omega_k + P_{k0} \ , \\[3ex]
 d_\pm = (d_1 \pm d_2) e^{\pm\Delta} \ ,  \qquad 
d_0 = d_1 \sin\!\delta_1 = - d_2 \sin\!\delta_2 \ , \qquad 
\displaystyle d_k = \frac{{\rm Im} \lambda_k}{|\lambda_k|^2}\\[3ex]
\omega_k:=\omega(\lambda_k) \ , \qquad  \omega (\lambda) = - 2\alpha\lambda^2-4\beta\lambda^3 \ ,
\ea \] \par \noindent
(by the way: changing the function $\omega(\lambda)$ we get other equations of a large hierarchy
of integrable systems, compare \cite{Ci-Nsol}), $\lambda_k$ are complex parameters, and, finally, 
$Q_{k0}$, $P_{k0}$, $\delta_1, \delta_2, \Delta$ are real constants.

The constants $Q_{k0}$, $P_{k0}$ are arbitrary but not very important, they just
fix the initial positions of the solitons.  The constants $\delta_1, \delta_2, \Delta$ depend 
on $\lambda_k$ as follows:
\be
  e^{\Delta + i\delta_1} :=
  \frac{(\lambda_1-\lambda_2) \overline{\lambda}_2}{(\lambda_1-\overline{\lambda}_2)\lambda_2} \ , \qquad
 e^{\Delta + i\delta_2} :=
  \frac{(\lambda_2-\lambda_1) \overline{\lambda}_1}{(\lambda_2-\overline{\lambda}_1)\lambda_1} \ .
\ee
Two complex parameters $\lambda_1$, $\lambda_2$ can be replaced by four real parameters
$a_1, a_2, b_1, b_2$ (where $\lambda_k = a_k + i b_k$), or, which is even more convenient, by $c_k, d_k$ expressed by:
\be \label{abcd}
c_k = \frac{a_k}{a_k^2 + b_k^2}
\ ,\quad d_k = \frac{b_k}{a_k^2 + b_k^2} \ , \quad
a_k = \frac{c_k}{c_k^2 + d_k^2}
\ ,\quad b_k = \frac{d_k}{c_k^2 + d_k^2} \ .
\ee
One  can also use the parameter $T_k$ (compare \cite{HBG}) responsible for the shape of 
the soliton: $T_k : = c_k/d_k \equiv a_k/b_k$.
The parameter $\Delta$ depends on $c_k, d_k$ as follows: 
\be  \label{Delta}
\Delta = \frac{1}{2} \ln
\frac{ (d_1 - d_2)^2 +
(c_1 - c_2)^2 }{ (d_1 + d_2)^2 + (c_1 - c_2)^2 }
\ee
The two-soliton formula \rf{N=2} is obtained
in the framework of the soliton surfaces approach
\cite{Ci-twosol,Ci-nos}. Its form is relatively simple (especially when
compared with other, equivalent, formulas, see \cite{FM,KI,S-6,LSW-vortex}) and
implies a lot of interesting consequences.
 Formulas for $N$-soliton solution
can be found for instance in \cite{Ci-Nsol} (see also
\cite{FM,KI,Ci-nos}). 

Let us first discuss the one-soliton solution
(in the case $\omega(\lambda)=-2\lambda^2$ found
by Hasimoto \cite{Ha}, extended for an arbitrary analytic
$\omega (\lambda)$ by Sym \cite{Sym}):
\be  \label{N=1}
{\bf r}_1 =  \left( \ba{r} 0 \\  0 \\
     -s \ea \right)  \
+ \frac{d_1}{\cosh\! Q_1} \left( \ba{c} \cos\! P_1 \\  \sin\!P_1 \\
        \sinh\! Q_1 \ea \right)  \ .
\ee \par \noindent

Physical characteristics of the single soliton solution
${\bf r}_{1}$
can be computed in the standard way. The amplitude of the Hasimoto vortex (i.e., its maximal 
distance from the $z$-axis) is given simply by $d_1$. The maximum of the wave
envelope ($Q_1=0$) performs a helical
movement and its $z$-coordinate ($z_{max1}$) moves at a constant
speed (group velocity $v^g_1$):
\be \label{group}
v^g_1 = {\dot z}_{max1} (t) =   \ \frac{{\rm Im} \omega_1}{{\rm Im} \lambda_1} =
- 4 (\alpha a_1 + \beta (3 a_1^2 - b_1^2)) \ .
\ee
The solitons can be classified  into
3 classes: lump solitons ($|a_1| > |b_1|$),
cusp solitons ($|a_1| = |b_1|$)  and loop solitons
($|a_1| < |b_1|$) \cite{KI}. The names
correspond to the shape of the curve $z = z(\rho)$
(where $\rho = \sqrt{x^2 + y^2}$). In experiments only the 
lump solitons were observed.
If the vortex filament has some loops, then the localized induction approximation
is not longer valid and these additional interactions
probably can destroy the loops. However, we remark that in the case of smoke rings 
(where such interactions are much more important)
the localized induction equation was successfully applied yielding some patterns 
which can be observed experimentally (\cite{CGrS}, compare \cite{Cho}).

The size of the lump soliton can be estimated by
its half-width. The half-width $D_1$ is defined as
a difference between the two values of $z$ corresponding to $Q_1$ satisfying $\cosh Q_1 = 2$:
\be  \label{halfwidth}
    D_1 =  d_1 \left( - \sqrt{3} + \left( 1 + \left( \frac{a_1}{b_1}
\right)^2  \right) \ln (2+\sqrt{3}) \right)
\ee
Another possibility is to estimate the wavelength of the soliton (i.e., the distance 
along the $z$-axis between  neighboring wavecrests) 
\be  \label{Lambda1}
 \Lambda_1 \approx  \frac{\pi}{a_1} \ . 
\ee
Not entering into technical details we mention that more precise computations 
lead to the result $\Lambda_1 = \pi/a_1 + d_1 ( \tanh Q'_1 - \tanh Q''_1)$,
where $Q'_1$ and $Q''_1$ depend on $s$ and $t$. The wavelength $\Lambda_1$ is not constant 
along the filament and depends on $t$. In experiments usually the central wavelength 
is measured. 
The wavelength $\Lambda_1$  equals $\pi/a_1$ with 
an accuracy $2 d_1/\Lambda_1$ which is less than $1/\pi$.
Usually, for $d_1 \ll \Lambda_1$, the accuracy is much better and,   
therefore, $\Lambda_1$ is treated as 
a constant value given by \rf{Lambda1}. Note that always 
$\Lambda_1 \ge 2 \pi d_1$. Obviously, the formulas \rf{group},\rf{halfwidth},\rf{Lambda1} 
are valid for the second soliton ($k=2$) as well.

The formula \rf{N=2} enables us to compute the asymptotic behavior of ${\bf r}_{2}$. 
Assuming $v^g_1 \neq v^g_2$  we consider the limit $Q_2
\rightarrow \pm\infty$ (or, more precisely, $\mbox{$\mid Q_1  \mid$} \ll
\mbox{$\mid Q_2  \mid$}$). Thus
\be \label{asymptota} {\bf r}_{2}
\ \ \stackrel{Q_2 \rightarrow \pm\infty}{\longrightarrow} \ \
 = \left( \ba{c}  0 \\ 0 \\ \pm d_2 - s  \ea   \right) +
  \frac{d_1}{\cosh\!(Q_1\pm\Delta)}
 \left( \ba{c} \cos\!(P_1\pm\delta_1) \\ \sin\!(P_1\pm\delta_1) \\
 \sinh\!(Q_1\pm\Delta)  \ea   \right)
\ee \par \noindent
The shape and the velocity of the soliton do
not change during the interaction. The only result of the
interaction is the  phase shift. In fact we have two phase shifts:
the shift $\Delta^{ph}_1$ along ${\bf e}_3$ axis and the shift of
the angular variable $P_1$. The first one is much more important
and can be measured in experiments. Analogical considerations 
are valid in the limit $Q_1 \rightarrow \pm\infty$.

Let us consider the case of the head-on collision of two
solitons (this case was observed in experiments \cite{MMH}).
We assume $v_1^g < 0$, $v_2^g > 0$ (head-on collision)
and also $b_k > 0$
(a technical assumption which can be made without loss of the
generality). We are going to compute the phase shift of the
first soliton (indexed by 1) after the interaction with the
second soliton.
If $t \rightarrow \pm \infty$, then $Q_2 \rightarrow \pm \infty$ and 
the $z$-coordinate of the envelope maximum for the first
soliton in asymptotic regions ($Q_2 \rightarrow \pm \infty$)
can be easily derived from \rf{asymptota}:
\be
  z_{max1}^{\pm} = v_1^g t + \frac{Q_{10}}{2 b_1} \pm d_2 \pm
\frac{\Delta}{2 b_1}
\ee
The phase shift $\Delta^{ph}_1 := (z_{max1}^+  - z_{max1}^-) {\rm sgn} (v^g_1)$ 
and, taking into account $v^g_1 < 0$, we have:
\be \label{phaseshift}
 \Delta^{ph}_1 = - 2 d_2 - \frac{\Delta}{b_1} =
 - 2 d_2 + \frac{c_1^2 + d_1^2}{2 d_1}
 \ln \frac{ (d_1 + d_2)^2 + (c_1 - c_2)^2 }{ (d_1 - d_2)^2 + (c_1 - c_2)^2 } \ .
\ee
Analogical considerations lead to the similar formula for the phase shift of the second soliton:
$\Delta^{ph}_2 = - 2 d_1 - \Delta/b_2$.

To compute the phase velocity of a soliton we have to determine positions of the individual
wave peaks. Considering the projection of the soliton on the 
$xz$-plane we have to solve (at least implicitly) the equation 
$\frac{  \partial }{ \partial   s}  \left( \frac{\cos P_k}{\cosh Q_k}    \right)   = 0$.
The resulting velocity of a wavecrest depends on its position $Q_1$: 
\be  \label{faz}
v^{ph}_k =  2  \alpha \frac{b_k^2 - a_k^2}{a_k} +
4 \beta (3 b_k^2 - a_k^2)  -
\frac{2 b_k^2 (\alpha + 4 \beta a_k) }{a_k \cosh^2 Q_k}
 \left( 3 - \frac{2 b_k^2}{ (a_k^2 + b_k^2) \cosh^2 Q_k }    \right)  .
\ee
In the analysis of solitons on a vortex filament the phase velocity
has been always assumed to be constant
sufficiently far from the envelope maximum
($Q_k \rightarrow \infty$). However in experiments usually the central region of the kink 
(i.e., $Q_k \approx 0$)
can be studied with the best accuracy. Anyway, the theoretical treatment of the phase velocity
is relatively difficult and have to be made with care. Experimental results 
concerning the phase velocity are also {\it ``rather crude''} \cite{MHR,MMH}.

Another quantity estimated from 
experimental observation is the rate of rotation of the filament maximum ($\Omega_k$),
namely $P_k (t)|_{Q_k=0} = \Omega_k t + \const$, where
\be
  \Omega_k = - 4 (\alpha + 4 \beta a_k) (a_k^2 + b_k^2) \ .
\ee

To compare the presented theory with experimental results
we recall that the two-soliton solution is parameterized by only
4 essential parameters $a_1, b_1, a_2, b_2$ (the parameters $Q_{k0}, P_{k0}$, related to
initial data, are not important for our analysis) and two more parameters
($\alpha, \beta$) characterize the physical model. In experiments
one can measure more or less precisely much more quantities: amplitudes ($d_k$), 
torsions, velocities ($v^g_k, v^{ph}_k$), widths ($D_k$, $\Lambda_k$), angular velocities ($\Omega_k$), phase shifts ($\Delta^{ph}_k$ and angular phase shifts) etc.

We choose $d_k$ and $D_k$ (or $\Lambda_k$) as fundamental quantities. They can be 
measured with reasonable accuracy. 
They are very convenient in theoretical analysis beaucause from \rf{halfwidth} we can 
express $c_k$ in terms of $d_k$ and $D_k$:
\be \label{ck}
 c_k = \pm d_k \left( \sqrt{ \frac{\frac{D_k}{d_k} + \sqrt{3}}{\ln (2 + \sqrt{3})}  - 1}
\right) \ .
\ee
and all other characteristics of the solution are expressed by $c_k, d_k$ 
(by virtue of \rf{abcd} we can use the variables $c_k, d_k$ instead of $a_k, b_k$).
From \rf{Lambda1} we can also express $c_k$ by $\Lambda_k$:
\be \label{celak}
   \frac{c_k}{d_k} = \frac{\Lambda_k}{2 \pi d_k} + 
\sqrt{ \left( \frac{\Lambda_k}{2 \pi d_k} \right)^2 - 1 } \ ,
\ee
where $c_k > 0$ is assumed. Combining \rf{ck} and \rf{celak} we get
\be  \label{deksi}
 D_k  = \left( 2.63  \xi_k \left( \xi_k + \sqrt{\xi_k^2 - 1} \right)  - 1.73 
\right) d_k \ , 
\ee
where  $\xi_k := \Lambda_k/(2 \pi d_k)$ (we recall that $\xi_k \geq 1$).  

According to the typical experimental situation we assume $d_1 \approx d_2$ and denote $d:=d_1$. Then
\be  \label{phs}
    \Delta^{ph}_k \approx ( -2 + 0.66 \tilde{\Delta} ) d + 0.38 \tilde{\Delta} D_k  
\ee
where $\tilde{\Delta}:= \ln (1 + 4 d^2 (c_1 - c_2)^{-2})$.

From \rf{deksi} it follows $D_k > \Lambda_k$ for $\Lambda_k/d_k > 10$ 
and $D_k > 2 \Lambda_k$ for $\Lambda_k/d_k > 17$.
Moreover, $\tilde{\Delta} > 1$  if $|c_1-c_2|< 1.53 d$, which can be estimated
as $|\Lambda_1-\Lambda_2| < 5 d$. Then $\Delta^{ph}_k \sim 0.4 D_k$ but this
relation can easily ``improved'' (e.g., $\Delta^{ph}_k \sim  D_k$) if solitons are
wider and more similar to each other. We arrive at the conclusion that 
head-on collision of lump solitons of similar size has to result in 
large phase shifts of both solitons.

The experimental observations of the soliton interactions have shown that 
{\it ``the forward advance (\ldots) is of the order of the central wavelength of the 
isolated kink wave''} (\cite{MMH}, compare Fig.\ 4). The analysis presented above is in perfect agreement with 
this observation. For $\tilde \Delta \approx 3$ we have a very interesting 
coincidence: the phase shift  is equal to the halfwidth of the soliton. 

We stress a very important point: the formulas \rf{phaseshift},\rf{phs} are valid for any 
$\alpha, \beta$. Therefore our conclusion does not depend on the parameters of the 
physical model (provided that we confine ourselves to the localized induction approximation 
\rf{axial}, of course), actually we can take even more general integrable equation described by an arbitrary function $\omega(\lambda)$.

We propose the following procedure of comparing the presented theory with experimental data.
First, taking $d_k$ and $D_k$ close to experimental values we use \rf{halfwidth} to compute $c_1, c_2$ (see \rf{ck}).
In order to get a large phase shift
we need $|c_1-c_2|$ to be rather small, therefore $c_1$ and $c_2$ should be of the same sign
(this sign is related to the sign of $\alpha$, which in turn depends on the
the direction of the vorticity field along the filament, we take $\alpha > 0$ \cite{MMH}). 
In any case we can use \rf{phaseshift} to choose right signs of $c_k$. 
Then, we use \rf{group} to compute $v^g_1$, $v^g_2$. Thus we have 
a system of two linear algebraic equations for $\alpha$ and $\beta$.
The system is invariant with respect to the transformation: $\alpha \rightarrow - \alpha$,
$c_k \rightarrow - c_k$ which means that an appropriate signs of $c_k$ yield $\alpha > 0$.
Finally, having all six parameters, we may compare other quantities (e.g., rotation rates
and phase velocities of the solitons) with experiment. Some quantities (e.g., phase shifts)
need only 4 parameters to be computed (see above).

 Konno and Ichikawa \cite{KI},
trying to fit their two-soliton solution to the experimental results of \cite{MMH},  
explained very well amplitudes and group velocities
of both solitons, to some extent their widths, and also the angular velocity and the phase 
velocity of one of the solitons (the corresponding experimental data for the second soliton 
were not reported in \cite{MMH}). 
We are able to explain (with reasonable accuracy) the same data but also both phase shifts.

Experimental data from \cite{MHR,MMH} (in units based on centimeter, second and radian) for the first soliton
read as follows:
$d_1= 0.49$, $v^g_1= - 33$, $\Lambda_1 = 6.8$, $\Omega_1 = - 2.4$, $v^{ph}_1 = -30$, the torsion $\tau_1 \equiv 2 a_1 = 0.89$. The second soliton is described with lower precision: $d_2 \sim d_1$,
$v^g_2 \approx - v^g_1$, $2 \Lambda_2 \sim \Lambda_1$. Finally, 
$\Delta^{ph}_k$ is ``of the same order'' as $\Lambda_k$. My rough estimate, based on  
\cite{MMH} (Fig.\ 4), is: $v^{g}_2=30$, $\Lambda_2 \sim 3.5$, $\Delta^{ph}_1 \sim  5$, 
$\Delta^{ph}_2 \sim 1.5$.

Performing the above procedure we get the following parameters: 
$a_1 = 0.46$, $b_1 = 0.11$, $a_2 = 0.78$, $b_2 = 0.28$. Then we assume
$v^g_1= - 33$, $v^{g}_2=30$ and finally we get: $d_1=0.49$, 
$d_2=0.41$,
$\Lambda_1 = 6.8$, $\Lambda_2 = 4.0$, 
$v^{ph}_1 = -30.5$, 
$\Omega_1 = -2.5$, 
$\Delta^{ph}_1 = 5.2$, $\Delta^{ph}_2 = 1.4$.

The agreement with experimental data is quite good (especially for 
the first soliton). Trying to get $d_2$ closer to the value $0.49$ 
we noticed an essential decrease of  $\Delta^{ph}_2$ or increase of $\Omega_1$.
Taking $d_2 = 0.41$ is a kind of compromise.

In this paper I revisited the Localized Induction Equation to show that,
in spite of popular views, that model explains interactions of solitons observed in experiments including the existence of the large phase shifts. 
A little bit surprising is that my result is equally true for the Localized Induction Equation with and without axial flow.

The theoretical investigation of the Localized Induction Equation, carried out
in many directions (compare \cite{FM,KI,Ci-nos,LP,SOK,Ni}), usually are
not easy to be used by experimentalists. 
I hope that the results of the present paper enable much better analysis of 
existing experimental
results (which seem to be published only fragmentarily) and  will prompt
further experimental work in this field. Of special interest should be 
any observations of the interaction of vortex solitons in the superfluid helium  because such vortices  are really thin which is assumed in the localized induction approximation.

{\bf Acknowledgements.} The research was partially supported by the Polish Committee for
Scientific Research (KBN grant No.\ 2 P03B 126 22).

\end{document}